\documentclass[proceedings]{JHEP3}

\PrHEP{PrHEP hep2001}                   
\conference{International Europhysics Conference on HEP}

\usepackage{epsfig}                   
\usepackage{amsmath}
\usepackage{amssymb}

\newcommand{\ltap}{\stackrel{<}{_\sim}}

\title{Flavor-Change with Ultra-Light Sbottom and Gluinos}

\author{Thomas Becher \\
        Stanford Linear Accelerator Center, Stanford University, Stanford CA 94309, USA }
\author{Stephan Braig and Matthias Neubert\\
        Newman Laboratory of Nuclear Studies, Cornell University, Ithaca, NY, 14853, USA }
\author{\speaker{Alexander L. Kagan}\\
        Department of Physics, University of Cincinnati, Cincinnati Ohio 45221, USA }

\abstract{Implications of a $2 - 5.5$ GeV
sbottom and $12 - 16$ GeV gluino masses for 
rare $B$ decay phenomenology are discussed.
An effective Hamiltonian is constructed in which the gluinos are integrated 
out and a $\tilde{b}$ squark remains among the light flavor degrees of freedom.
Restrictive constraints come from 
$b \to s \gamma $ and $b \to sg $, but they allow 
a substantially enhanced inclusive $b$ decay rate
into charmless hadronic final states, and ${\cal O}(10\%)$ direct CP asymmetries  
in $B \to X_s \gamma $ and $B^\pm \to K^0 \pi^\pm  $ decays,  
which are an order of magnitude larger than in the Standard Model.
New contributions to $B_s $ mixing are negligible but significant
effects in $B_d $ mixing may be possible.}

\begin{document}

\section{Introduction and Motivation}

The measured $b$ quark production cross section at 
hadron colliders has persistently exceeded NLO QCD predictions by 
factors of 2.  Rather than attributing 
this discrepancy to additional QCD contributions, e.g., 
those arising at NNLO, it is interesting to ask whether New Physics could be 
responsible. In \cite{berger} Berger {\it{et al.}} have shown 
that gluino pair production, followed by decay of each
gluino to a bottom-sbottom pair can account 
for the missing rate if the $\tilde{g}$ and light $\tilde{b}$ masses lie in the ranges
$m_{\tilde{g}} \cong 12-16$ GeV and $m_{\tilde{b}} \cong 2-5.5$
GeV, respectively.  They have further observed that a light $\tilde{b}$ squark 
could have evaded direct detection.  For example,  
the additinal contribution to $R_{had}$ at large $\sqrt{s}$ would only 
be $\cong 2\%$, and hence difficult to disentangle.  In the resonance region, e.g.,
$\sqrt{s} \sim 5 - 8$ GeV, a light $\tilde{b}$ squark may resolve a long standing discrepancy
in $R_{had}$ between the MARK I and Crystall Ball Collaborations \cite{srivastava}. 

There are important $Z$-pole constraints on the light
$\tilde{b}$, $\tilde g$ scenario \cite{wagner}.  Most importantly, 
the light sbottom's coupling to the
$Z$ must be suppressed. The light and heavy 
sbottom mass eigenstates $\tilde{b} $ and
$\tilde{b}_H$, rspectively, are admixtures of the 
left-handed and right handed bottom quark
superpartners, $\tilde{b}_L$ and $\tilde{b}_R$, characterized by a mixing angle
$\theta_{\tilde{b}}$: $\tilde{b} = \cos\theta_{\tilde{b}} \tilde{b}_R + \sin\theta_{\tilde{b}}
\tilde{b}_R $.
The coupling of $\tilde{b} $ to the $Z$ is proportional to 
$- \sin^2 \theta_{\tilde{b}}/2  + \sin^2 \theta_W /3 $,
and vanishes at tree-level if $\sin\theta_{\tilde{b}} \cong .38$. This
implies that the light sbottom must be predominantly `right-handed{'}.
An overall fit to $Z$ pole observables in \cite{wagner}, 
which however only considered the impact of light sbottoms
at tree-level, 
finds a slight improvement over the Standard Model fit
for $\sin\theta_{\tilde{b}}$ in the range [.3,.45]. 
A recent study \cite{cao} finds very restrictive contraints on $\sin\theta_{\tilde{b}}$ 
from light $\tilde{b}$-$\tilde{g}$ loop
contributions to $R_b$, however it does not include a fit to 
all $Z$-pole observables and does not take into account other potentially important
one-loop supersymmetric contributions.

The presence of a light sbottom and light gluinos
alters the running of $\alpha_s$. In the following we will root 
the evolution of $\alpha_s (\mu) $ at low scales 
using determinations from $\tau $
decays and deep
inelastic scattering 
at scales $\mu \ltap 5 $ GeV, which would be
unaffected by the new particles.
We consider $\alpha_s (m_b ) \cong
.19 - .22$. The running of $\alpha_s$ at larger scales
is slower than in the Standard Model,
e.g., $\alpha_s (M_Z) \approx .121 - .133$, but the lower range
of predicted values is compatible with experiment.
The comparison is illustrated in Figure 1, where the running 
of $\alpha_s$ in the two cases is confronted with a compilation 
of measurements collected in
\cite{bethke}.  A new LEP2 average at $\sqrt{s} = 206$
GeV is also included \cite{flagmeyer}.  A fit to the data 
appears to be only marginally better in the Standard
Model.  However, it should be noted that in the supersymmetric 
scenario the high energy observables used to determine
$\alpha_s$ have not been corrected for new 
contributions due to the light $\tilde{b}$ squark and gluinos.

\FIGURE{
\epsfxsize=0.5\textwidth
\epsfbox{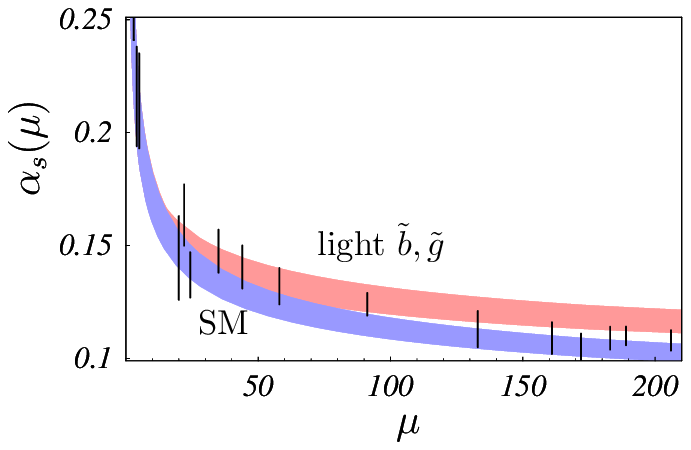} 
\caption{Runnnig of $\alpha_s$ in the Standard Model (blue band),
and with $m_{\tilde{b}} = m_b$, $m_{\tilde{g}} = 15$ GeV (pink band),
for $ .19 < \alpha_s (m_b)< .22$. The data points are from a compilation
of experimental determinations, see text. \label{fig:Fig1} } }

With regards to potential light $\tilde{b}$ decay modes, a null CLEO search for 
semileptonic decays \cite{CLEOsemilep} implies that the branching ratios for the 
decay modes
$\tilde{b} \to c \ell $, induced by $R$-parity breaking couplings, or
$\tilde{b} \to c \ell \chi^0 $, where $\chi^0$ is an ultra-light neutralino,
must be highly suppressed.  However, the light $\tilde{b}$ squark is allowed to
decay promptly via hadronic $R$-parity breaking couplings
in the modes $\tilde{b} \to \bar{c} \bar{q},~\bar{u} \bar{q}$, $q=u,s$.
Alternatively, it could be long lived, forming $\tilde{b}$-hadrons. 

If we are to take the possibility of a light $\tilde{b}$ squark and light gluinos seriously then
the theoretical study of their impact must be extended to 
$b$ decay phenomenology, which is currently undergoing intensive scrutiny at
the $B $ factories.  In this talk we report on work in progress in this direction.
New sources of flavor violation could arise via supersymmetric
$s- \tilde{b} -\tilde{g}$ and $d- \tilde{b} -\tilde{g}$ `Yukawa{'} couplings.
The overall scale of supersymmetric flavor violating interactions
originating from gluino exchange is set 
by the factor $g_s^2 /m^2_{\tilde{g}}$, which is
much larger than the corresponding factor $G_F /\sqrt{2} \sim g_W^2 /M_W^2$
for weak decays in the Standard Model.
Consequently, the new
flavor-changing couplings must be
much smaller than the corresponding CKM mixing angles.
We will find that these couplings must 
be less than $10^{-4}$, 
in order to satisfy constraints
coming from virtual gluino-sbottom loop contributions
to $b \to s \gamma $, $b \to sg$,
and $B \to K \pi$. Conversely, even with tiny flavor-changing couplings 
large deviations from Standard Model predictions are possible.   
This is of particular
interest from a  model-building point of view.

The phenomenology of this scenario
will depend on whether or not light $\tilde{b}$ squarks can be 
pair-produced in $b$ decays.  If they are too heavy
then they only give rise to virtual effects, which we discuss here.
However, if $\tilde{b} $ squarks are light enough to be pair produced,
new unconventional decay channels would be opened up for $B$
mesons and beauty baryons. Potentially interesting consequences
of such decays are briefly mentioned in the Conclusion.

\section{The low energy effective Hamiltonian}

Flavor-changing processes in the light sbottom-gluino model 
are most transparently
described by means of an effective low-energy Hamiltonian in which the effects of the 
`heavy{'} gluino fields are integrated out.  
The light degrees of freedom 
are the light quarks $u, d, s, c, b$,
the photon and gluons, as well as the light $\tilde{b} $ squark.
An expansion of the low-energy Lagrangian in powers of $1 /m_{\tilde{g}}$ is justified
for $m_{\tilde{g}} \cong 12-16$ GeV.  For rare $B$
decays we have checked that to good approximation it is 
sufficient to work to leading order in this expansion.

To parametrize the flavor-violating couplings entering the Hamiltonian,
let $\tilde{d}_i $, $i=1,..,6$ denote the down squark mass
eigenstates, and $\tilde{d}_L^I, \tilde{d}_R^I$, $I=1,2,3$ denote
the interaction eigenstates (superpartners of the left-handed and right-handed
down quarks).  We write in the usual way \cite{haberkane}
\begin{equation}
\tilde{d}_L^I = (\Gamma^L )_{Ii}^\dag \tilde{d}_i,~~~~
\tilde{d}_R^I = (\Gamma^R )_{Ii}^\dag \tilde{d}_i,
\label{eq:Gammadef}
\end{equation}
and identify $\tilde{d}_3$ with the light sbottom.
The rest of the squark masses are taken to be of order the generic supersymmetry breaking mass,
$M_{SUSY} \ltap 1$ TeV.
The new flavor-violating effects arising from light $\tilde{b}$ and
$\tilde{g}$ exchange can be parametrized by the dimensionless quantities, 
\begin{equation}
\epsilon_{i3}^{AB} \equiv (\Gamma^A )_{i3}^\dag (\Gamma^B )_{33},
~~~i=1,2; ~~~ A,B = L~ \rm{or}~ R.
\end{equation}
Note that in general they can be complex, which would lead to new CP violating 
effects.
In terms of the sbottom sector mixing angle introduced earlier,
$\Gamma_{33}^R = \cos\theta_{\tilde{b}}$ and $\Gamma_{33}^L = \sin\theta_{\tilde{b}}$,
implying the following relations,
\begin{equation}
\epsilon^{LL}_{23} = \epsilon_{23}^{LR}/ \cot\theta_{\tilde{b}},~~~~
\epsilon^{RR}_{23} = \epsilon_{23}^{RL}  \cot\theta_{\tilde{b}},
\label{eq:epsrelns}
\end{equation}
and similarly for the $\epsilon^{AB}_{13}${'}s.

In general, new contributions to the $\Delta B = 1$ effective Hamiltonian can be written 
as
\begin{equation}
{\cal H}_{\Delta \! B\!=\!1} = \frac{4 \pi \alpha_s }{m_{\tilde{g}}^2} \sum 
\left( C_i^T \!(\! \epsilon_{23}^{AB}\! ) T_i + C_i \! (\! \epsilon_{23}^{AB} \! )
{\cal O}_i \right),
\label{eq:hamiltonian}
\end{equation}
where  the dependence of the Wilson coefficients
on the flavor-violation parameters $\epsilon_{23}^{AB}$ has been indicated.  We briefly
describe the operators which arise below.
For brevity we explicitly include only the $\Delta S = 1$
operators. The $\Delta S = 0$ operators follow by substituting
$s \to d$ everywhere.  
Their Wilson coefficients follow from the substitutions 
$\epsilon_{23}^{AB} \to \epsilon_{13}^{AB}$. Color indices are suppressed throughout. 
The $T_i $ arise from tree-level matching
of the full theory onto the effective theory at scales $\mu \sim m_{\tilde{g}}$:
\begin{itemize}
\item Eight $T_i$ operators are present at order $1/m_{\tilde{g}}$, four of the form 
$\bar{s} (1 \pm \gamma_5 ) b \tilde{b}^*  \tilde{b}$ with strenghths
depending linearly on $\epsilon^{LR} _{23}$ or $\epsilon^{RL} _{23}$,
and four of the form $\bar{s^c}  (1\pm \gamma_5 ) b \tilde{b}^*  \tilde{b}^*$  
with similar dependence on $\epsilon^{LL} _{23}$ or $\epsilon^{RR} _{23}$.  The
latter could mediate rare $B$ decys to `wrong-sign{'} kaons.
\item Eight $T_i$ operators arise at order $1/m_{\tilde{g}}^2 $, and can therefore be neglected
to good approximation. 
Four are of the form $\bar{s} \gamma_\mu (1\pm \gamma_5 ) b \tilde{b}^* D^\mu \tilde{b}$, 
with strenghths depending linearly on $\epsilon^{LL} _{23}$ or $\epsilon^{RR} _{23}$, and
four are of the form
$\bar{s^c} \gamma_\mu (1\pm \gamma_5 ) b \tilde{b}^* D^\mu \tilde{b}^*$,
with similar dependence on $\epsilon^{LR} _{23}$ or $\epsilon^{RL} _{23}$.
\end{itemize}
\indent The Wilson coefficients for the `one-loop{'} operators ${\cal O}_i$ are obtained by
computing the full theory amplitudes due to light 
$\tilde{b}$-$\tilde{g}$ loops and subtracting the corresponding 
light $\tilde{b}$-loop contributions of the $T_i$.
\begin{itemize}
\item Four such operators are present at order $1/m_{\tilde{g}}$:
Two electromagnetic dipole operators (Standard Model and opposite-chirality) which mediate $b \to s
\gamma$ decays, of the form $\bar{s} \sigma_{\mu\nu} (1 \pm \gamma_5 ) e F^{\mu\nu} b$,
and two chromomagnetic
dipole operators (Standard Model and opposite-chirality) which mediate $b \to sg $ decays,
of the form $\bar{s} \sigma_{\mu\nu} (1 \pm \gamma_5 ) g_s G^{\mu\nu} b$.  
\item Eight four-quark QCD penguin operators arise at order $1/m_{\tilde{g}}^2 $:
four are of the same form as in the Standard Model and four have the opposite chirality, i.e., 
$ \bar{s} \gamma_{\mu} (1 - \gamma_5) b \sum_q  \bar{q} \gamma^{\mu} (1 \pm \gamma_5)
q $ and $ \bar{s} \gamma_{\mu} (1 + \gamma_5) b \sum_q  \bar{q} \gamma^{\mu} (1 \pm \gamma_5)
q $, respectively. Their effects can
be neglected compared to those of the chromomagnetic dipole operators.
\end{itemize}
New contributions to the Standard Model chirality dipole operator
Wilson coefficients depend linearly on $\epsilon_{23}^{LR}$ 
(at leading order in $1/m_{\tilde{g}}$), whereas the opposite-chirality coefficients depend
on $\epsilon_{23}^{RL}$. The Standard Model and opposite-chirality 
QCD penguin Wilson coefficients depend linearly on $\epsilon_{23}^{LL}$ and 
$\epsilon_{23}^{RR}$, respectively.  Although the latter 
can be neglected, it is interesting to note that
the sbottom mixing angle $\theta_{\tilde{b}}$ in Eq. (\ref{eq:epsrelns})
fixes the ratios of (Standard Model or opposite-chirality) dipole operator
to QCD penguin operator Wilson coefficients.

Finally, eight $\Delta B = 2 $, $\Delta S = 2$ operators which can mediate $B_s$ mixing are
present at order $1/m_{\tilde{g}}^2$ after one-loop matching of the full theory 
sbottom-gluino box graphs
onto the effective theory.  They are of the form 
$\bar{s} \gamma_\mu (1 \pm \gamma_5 ) b  \bar{s} \gamma^\mu (1 \pm \gamma_5 ) b$, 
$\bar{s} \gamma_\mu (1 \pm \gamma_5 ) b  \bar{s} \gamma^\mu (1 \mp \gamma_5 ) b$,
and $\bar{s}  (1 - \gamma_5 ) b  \bar{s} (1 - \gamma_5 ) b$, 
(as ususal color indices have been suppressed). Analogous operators mediating $B_d$ mixing 
are obtained by substituting $s\to d$ everywhere.
We can write the effective $\Delta B = 2$ Hamiltonian in the form 
${\cal H}_{\Delta\! B\! =\! 2}  = \alpha^2_s / m_{\tilde{g}}^2  \sum 
D_i  {\cal Q}_i $.
The Wilson coefficients $D_i$ for the $\Delta S = 2$ and $\Delta S = 0$ operators depend
quadratically on the $\epsilon_{23}^{AB}$ and $\epsilon_{13}^{AB}$, respectively .

\section{Rare $B$ decays and $B$ mixing}

We begin with a discussion of constraints coming from $B \to X_s \gamma $ and $B \to X_{sg}$
decays.   Our strategy for describing $B \to X_s \gamma $ decays is to perform a partial 
next-to-leading order (NLO)
analysis: The Standard Model contributions are included fully at NLO, as in 
\cite{greub,misiak,kaganneubertanat}, since we know that
the NLO corrections, particularly
those due to the $\bar{c}_L \gamma_\mu b_L \bar{s}_L \gamma^\mu c_L $
current-current operator, are substantial. However, the 
new SUSY contributions are accounted for at leading-order (LO). 
We confront theoretical predictions
with the new CLEO branching ratio measurement \cite{CLEObsgamma}, 
$\rm{BR}(B\to X_s \gamma) = (3.06 \pm .41 \pm .26) \times 10^{-4} $, obtained for
$E_\gamma > 2$ GeV.  If the 
chromomagnetic $b \to sg$ dipole operators are signifcantly
enhanced, the shape of the photon energy spectrum is
modified by new soft contributions from photon bremsstrahlung.
Due to this possibility, we compare branching-ratio predictions
directly with the CLEO measurement for $E_\gamma > 2 $ GeV, using shape function
convolutions for the energy spectrum \cite{kaganneubertanat}.
For simplicity, we limit our discussion to constraints on new contributions to the 
Standard Model chirality dipole operators, which only depend on 
$\epsilon_{23}^{LR}$ at leading order in $1/m_{\tilde{g}}$.  
To very good approximation we can describe these processes
using a truncated operator basis, ignoring the QCD penguin operators.

We use the parametrization
$\epsilon_{23}^{LR} = |\epsilon_{23}^{LR} | e^{i \theta_{LR}}$, and exhibit constraints 
as contours in the ($|\epsilon_{23}^{LR} |$, $\theta_{LR}$) plane. 
In Figure 2a CLEO $\pm 1 \sigma$ contours are drawn for 
$\rm{BR}(B\to X_s \gamma)_{E_\gamma > 2}$ (CP-averaged). 
The ratio $z = m_c /m_b $ entering the charm-loop $b \to s \gamma$ matrix element 
of the current-current operator has been allowed 
to vary in the range [.22,.29]. Values near .22 are obtained by using
the $\overline{\rm MS} $ mass, $m_c (m_b )$ 
\cite{misiakmass}. Elsewhere, $z = .29$ is used. 
Absence of significant tuning
evidently requires $|\epsilon_{23}^{LR} | < 1 \times 10^{-4}$. 
This conclusion is further reinforced by the contours for 
$\rm{BR}(B\to X_{sg})$ shown in Figure 2b. The CLEO upper-bound\footnote{Using more recent
inclusive charmonium and charmed baryon yields gives an upper bound of 9\%.}
of $6.8\%$ (90\% c.l.) \cite{CLEObsglue} significantly reduces the 
allowed region, so that $|\epsilon_{23}^{LR} | \ltap 5 \times 10^{-5}$.
However, a sizable new weak phase $\theta_{23}^{LR}$ is allowed.
In Figure 2c we have drawn contours for the direct CP asymmetry
$A_{CP} (B \to X_s \gamma)$ making use of formulae in \cite{kaganneubertCP}, and have
included the 
CLEO 90\% c.l. upper and lower bounds of +10\% and -27\%, respectively \cite{CLEOAcp}.
Finally, in Figure 2d we show the allowed region which survives the
three constraints.  Comparison with the contours 
for $A_{CP} $ and $\rm{BR}(B\to X_{sg})$ implies that 
$A_{CP} \sim 10\% $ is possible, as is a significantly enhanced
$b \to sg $ branching ratio of 5 - 10 \%.
Recall that in the Standard Model $A_{CP} \sim 1\%$ and 
$\rm{BR}(B\to X_{no~charm}) \sim 1\% $.

We have not yet taken into account potentially important 
two-loop ${\cal O}(1/m_{\tilde{g}} )$ contributions
to the LO anomalous dimension matrix from
mixing of the current-current operators $T_i$ into
the dipole operators.
This work is currently in progress.
There may also be important contributions arising at NLO from  
$b \to s \gamma$ matrix elements of the operators $T_i$.
These corrections will amount to a `$k$-factor{'} rescaling of the
the $|\epsilon_{23}^{LR}|$ axes in Figure 2.  
It should also be noted that the theoretical predictions for $A_{CP}$ suffer from
large renormalization scale dependence. 
Therefore, at this stage we regard the constraints shown in Figure 2 as
illustrative. Nevertheless, our conclusions hold qualitatively.

\FIGURE{
\epsfxsize=0.7\textwidth
\epsfbox{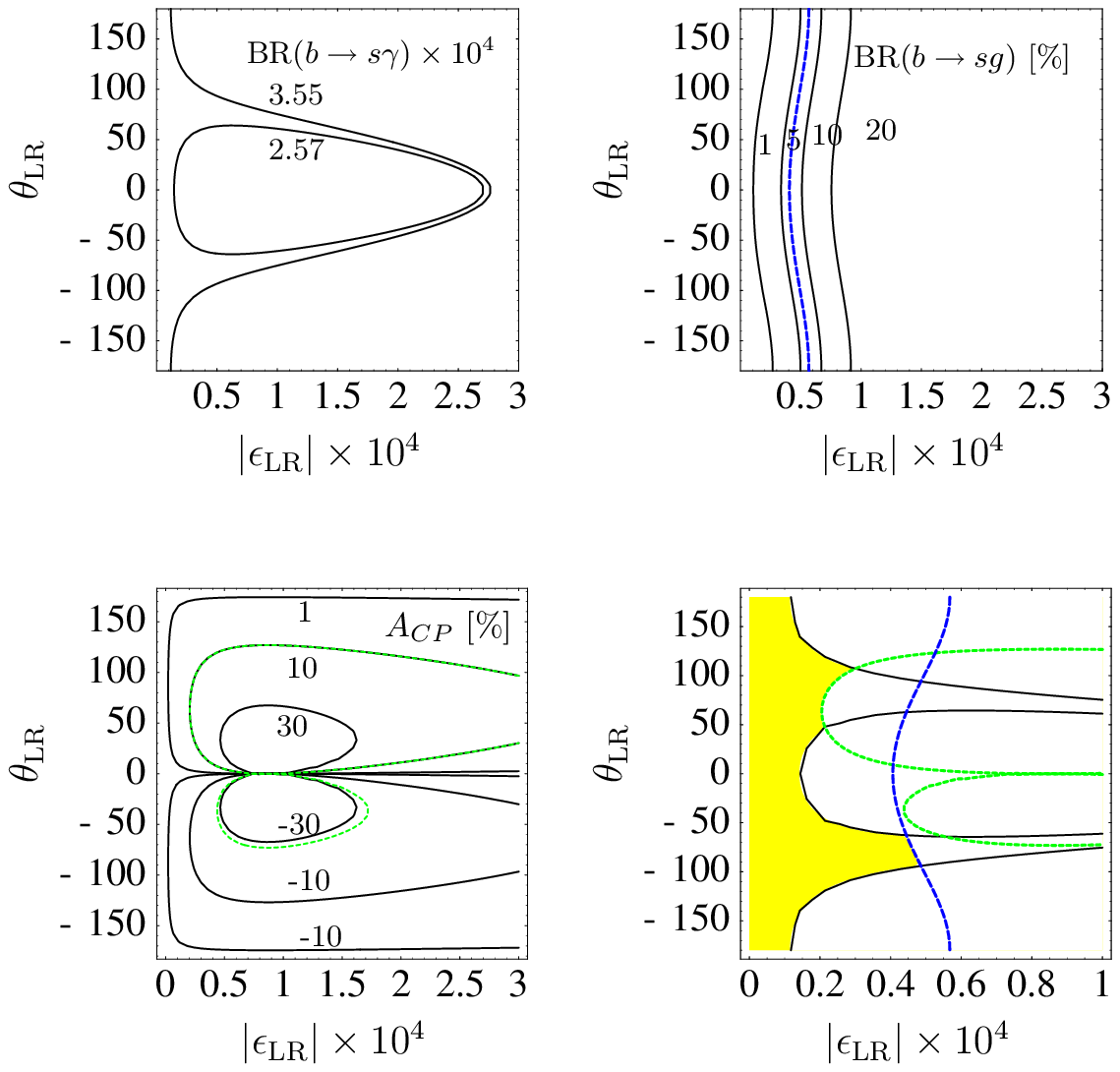} 
\caption{(a) $\pm 1 \sigma $ contours for the CLEO $\rm{BR}(B\to X_s \gamma)_{E_\gamma > 2}$
measurement in the ($|\epsilon_{23}^{LR} |$, $\theta_{LR}$) plane. (b) Contours for 
$\rm{BR}(B\to X_{sg})= 1,5,10,20 \%$.  Also shown (blue-dashed) is the CLEO
upper bound of 6.8\%. (c) Contours for $A_{CP} (B \to X_s \gamma) = \pm 1, \pm10,
\pm 30 \% $. Also shown (green-dotted) are the 90\% c.l. upper and lower
limits from CLEO. (d) Combination of the previous bounds highligting the allowed region
(shaded yellow) in the ($|\epsilon_{23}^{LR} |$, $\theta_{LR}$) plane.
$\alpha_s (M_z) = .125$ $(\alpha_s (m_b) = .20 )$, $\mu = m_b = 4.8$ GeV.
\label{fig:Fig2} }}

Rare $B \to K \pi $ decays are also described to very good approximation
at leading order in $1/m_{\tilde{g}} $, so we need only keep new contributions to the
chromomagnetic dipole operators.
Only Standard Model chirality operators are considerd so 
that constraints can again be simply exhibited in the ($|\epsilon_{23}^{LR} |$, $\theta_{LR}$) plane.
All contributions to the amplitudes have been evaluated using the QCD factorization
approach of Beneke {\it{et al.}} (BBNS) \cite{bbns}, taking their default 
values for the various hadronic and CKM input parameters.
Here we report on results for the decays $B^\pm \to K^0 \pi^\pm $, 
which have little sensitivity to the CKM weak phase $\gamma$ 
and which therefore are predicted to have a 
very small direct CP asymmetry in the Standard Model.

In Figure 3a, $\pm 1 \sigma $ contours have been drawn for the World Average
branching ratios quoted in \cite{bbns}, 
${\rm{BR}}(B^\pm \to K^0 \pi^\pm )_{\rm{W\!~Avg}} = (17.2 \pm 2.5)\times 10^{-6}$.  A 
second set of contours has been added for 1.35 $\times (+ 1\sigma $ BR) and 
.65 $\times (-1\sigma $ BR),
which is intended to take 
into account a typical $\pm 35\% $ uncertainty in BBNS
branching ratio predictions when varying over all input parameters.
By superimposing the allowed region from $B \to X_s \gamma,~X_{sg}$ in Figure 2d,
the $B \to K \pi $ constraints are seen to be comparable but less restrictive.
Contours for $A_{CP} ( B^\pm \to 
K^0 \pi^\pm )$ obtained using the BBNS approach are shown in Figure 3b. 
Comparison with the allowed region
indicates that large direct CP asymmetries of order 10\% are possible,
to be compared with $\sim 1\%$ in the Standard Model.  
A similar result is obtained for $A_{CP} (B^\pm
\to \phi K^\pm)$.  
Although the theoretical uncertainties for $A_{CP} (K \pi )$ are large \cite{bbns},
and 
two-loop mixing of the $T_i$ into the chromomagnetic dipole operators
has not been taken into account,
such added effects again will not
change our conclusions qualitatively.

\FIGURE{
\epsfxsize=0.7\textwidth
\epsfbox{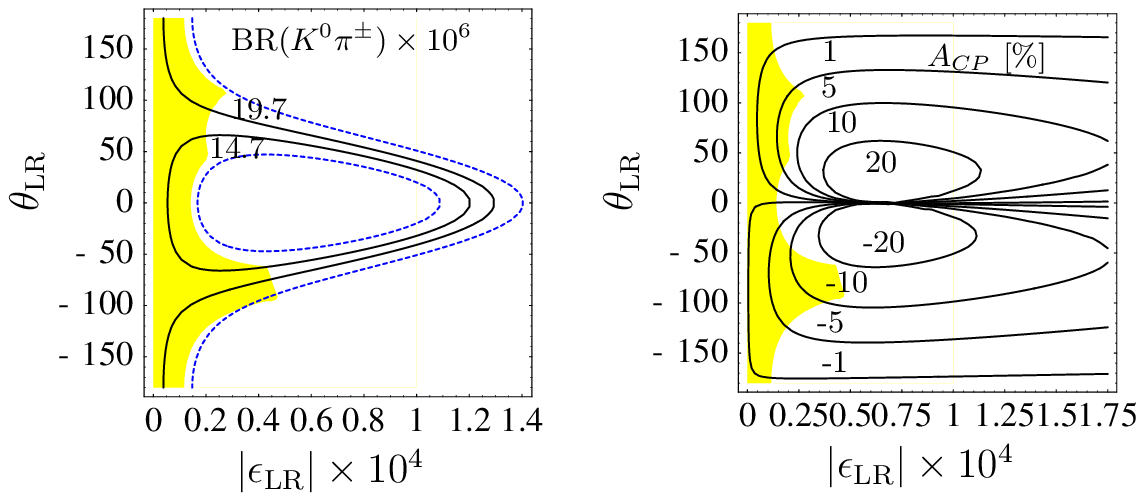} 
\caption{(a) $\pm 1 \sigma $ contours for 
${\rm{BR}}(B^\pm \to K^0 \pi^\pm )_{world~avg}$ in the 
($|\epsilon_{23}^{LR} |$, $\theta_{LR}$) plane,
using default BBNS inputs (solid-lines), and assuming a
$\pm 35\%$ uncertainty (blue-dashed), see text.  The allowed region from radiative decays
is shaded yellow. (b) Contours for $A_{CP} (K^0 \pi^\pm) = \pm 1,\pm5,\pm10,
\pm 20$ \%. $\alpha_s (M_z) = .125$, $\mu = m_b = 4.2$ GeV.
\label{fig:Fig3} }}

We conclude this section with a brief discussion of $B$ mixing constraints. 
Contributions of the $\Delta B = 2$ operators 
to $\Delta M_{B_d}$ involve several different combinations of the $\epsilon_{13}^{AB}${'}s.
Requiring that the contribution of each operator
by itself should not exceed the measured value of $\Delta M_{B_d}$, 
and using the vacuum saturation approximation, we
find for example that $\sqrt{{\rm Re}[\epsilon_{13}^{LR} \epsilon_{13}^{LR} ]} < (1-4) \times
10^{-4}$.
This is not as restrictive as the bounds obtained
from radiative $b \to d \gamma $ decays.
Given that the CLEO measurement of 
the inclusive radiative branching ratio actually corresponds to the sum of
$b \to s\gamma$ and $b \to d \gamma$, 
and that $\Gamma (B \to \rho\gamma)/\Gamma (B \to K^* \gamma ) < .19$ (90\% c.l.) 
\cite{EXPdgammasgamma}
, the bound on $|\epsilon_{13}^{LR}|$ from radiative 
$B $ decays is at least as stringent as the bound on $|\epsilon_{23}^{LR}|$.
Therefore, new supersymmetric contributions probably could not account for 
the bulk of $\Delta M_{B_d}$, but they may significantly modify
the CP violating mixing phase.
Finally, bounds on $\epsilon_{23}^{AB} $ from radiative $B$ decays
imply that new supersymmetric contributions to $B_s $ mixing must be negligible.

\section{Conclusion}

It has been pointed out that new supersymmetric contributions to $b$ quark production at
hadron colliders can account for the
long-standing discrepancy between the measured and NLO QCD cross sections
if there is a light $\tilde{b}$ squark with mass in the range 2 - 5.5 GeV, and 
if the gluinos have mass in the range 12 - 16 GeV \cite{berger}.
In this talk we have explored the phenomenology of rare $B$ decays in such a scenario,
and have found very restrictive constraints on the flavor-violation parameters
controlling supersymmetric contributions to $b \to s $ and $b \to d$ transitions, 
namely $\epsilon_{23}^{AB},~\epsilon_{13}^{AB} \ltap \rm{a~few} \times 10^{-5}$.
This implies that certain off-diagonal down squark mass matrix entries 
must be similarly suppressed compared to the generic squark mass squared.
However, interesting New Physics effects are possible. Among these are
an enhanced $B \to X_{no~charm}$
rate, which may help explain the low $b$ semileptonic branching ratio and 
charm mulitplicity, and ${\cal O}(10\%)$ direct CP asymmetries
in $B \to X_s \gamma $ and $B^\pm \to K^0 \pi^\pm $ decays, to be compared with 
Standard Model CP asymmetries of order 1\%.
The most restrictive constraints are due to $B\to X_s \gamma$ 
and $B \to X_{sg}$. Less restrictive constraints 
follow from $B_d$ mixing, and new contributions to $B_s $ mixing must be negligible.
We have not considered the potential consequences of $\tilde{b}$ pair production.  
The new decay
modes $b \to s \tilde{b}^* \tilde{b} $ and $b \to \bar{s} \tilde{b} \tilde{b}$
which become accesible when the
$\tilde{b}$ is sufficiently light 
would affect the decay widths of $B$ mesons and $\Lambda_b$ baryons
differently, and hence could potentially explain the anomaly of 
the low $\Lambda_b$ lifetime. A significant increase in the $\Gamma_{B_d} -
\Gamma_{\overline{B}_d}$ lifetime difference may also be possible.
We will report on this interesting class of effects elsewhere.

\end{document}